\documentclass[aps,prl,superscriptaddress,twocolumn,showpacs,titlepage=false]{revtex4-1}
\usepackage{graphicx}
\usepackage{amsmath}
\usepackage{amssymb}
\usepackage[breaklinks=true,colorlinks=true,linkcolor=blue,urlcolor=blue,citecolor=blue]{hyperref}

\def\ket#1{{\left| #1 \right\rangle}}

\begin{document}

\title{Practical quantum metrology with large precision gains in the low photon number regime}
\author{P. A. Knott}
	\email{P.Knott@Sussex.ac.uk}
	\affiliation{Department of Physics and Astronomy, University of Sussex, Brighton BN1 9QH, United Kingdom}
\author{T. J. Proctor}
	\affiliation{School of Physics and Astronomy, University of Leeds, Leeds LS2 9JT, United Kingdom}
\affiliation{Berkeley Quantum Information and Computation Center, University of California, Berkeley, CA 94720, USA}
\author{A. J. Hayes}
	\affiliation{Department of Physics and Astronomy, University of Sussex, Brighton BN1 9QH, United Kingdom}
\author{J. P. Cooling}
	\affiliation{Department of Physics and Astronomy, University of Sussex, Brighton BN1 9QH, United Kingdom}
\author{J. A. Dunningham}
	\affiliation{Department of Physics and Astronomy, University of Sussex, Brighton BN1 9QH, United Kingdom}


\date{\today}

\begin{abstract}
Quantum metrology exploits quantum correlations to make precise measurements with limited particle numbers. By utilizing inter- and intra- mode correlations in an optical interferometer, we find a state that combines entanglement and squeezing to give a 7-fold enhancement in the quantum Fisher information (QFI) -- a metric related to the precision -- over the shot noise limit, for low photon numbers. Motivated by practicality we then look at the squeezed cat-state, which has recently been made experimentally, and shows further precision gains over the shot noise limit and a 3-fold improvement in the QFI over the optimal Gaussian state. We present a conceptually simple measurement scheme that saturates the QFI, and we demonstrate a robustness to loss for small photon numbers. The squeezed cat-state can therefore give a significant precision enhancement in optical quantum metrology in practical and realistic conditions.
\end{abstract}
\maketitle



\section{Introduction}

Optical quantum metrology utilizes quantum mechanical correlations to make high precision phase measurements with a significantly lower particle flux than would be required by classical systems. This is a crucial requirement for many applications such as biological sensing, where disturbing the system can damage the sample \cite{wolfgramm2013entanglement,taylor2013biological}, or gravitational wave detection, which suffers from the effects of radiation pressure and mirror distortion if the photon flux is too high \cite{punturo2010third,purdy2013observation}. Squeezed states of light have shown much promise for quantum-enhanced metrology beyond the classical shot noise limit (SNL), and since the seminal proposal of Caves \cite{caves1981quantum} significant progress has been made in exploiting the potential of such states \cite{pezze2008mach,sahota2015quantum,ono2010effects,seshadreesan2011parity}. As a result the effectiveness of squeezing in quantum metrology has been demonstrated experimentally \cite{xiao1987precision}, and a squeezed vacuum is now routinely injected into the dark port of gravitational wave detectors to improve their measurements \cite{vahlbruch2010geo,aasi2013enhanced,grote2013first}.

\begin{figure}[t]
\centering
\includegraphics[scale=0.62]{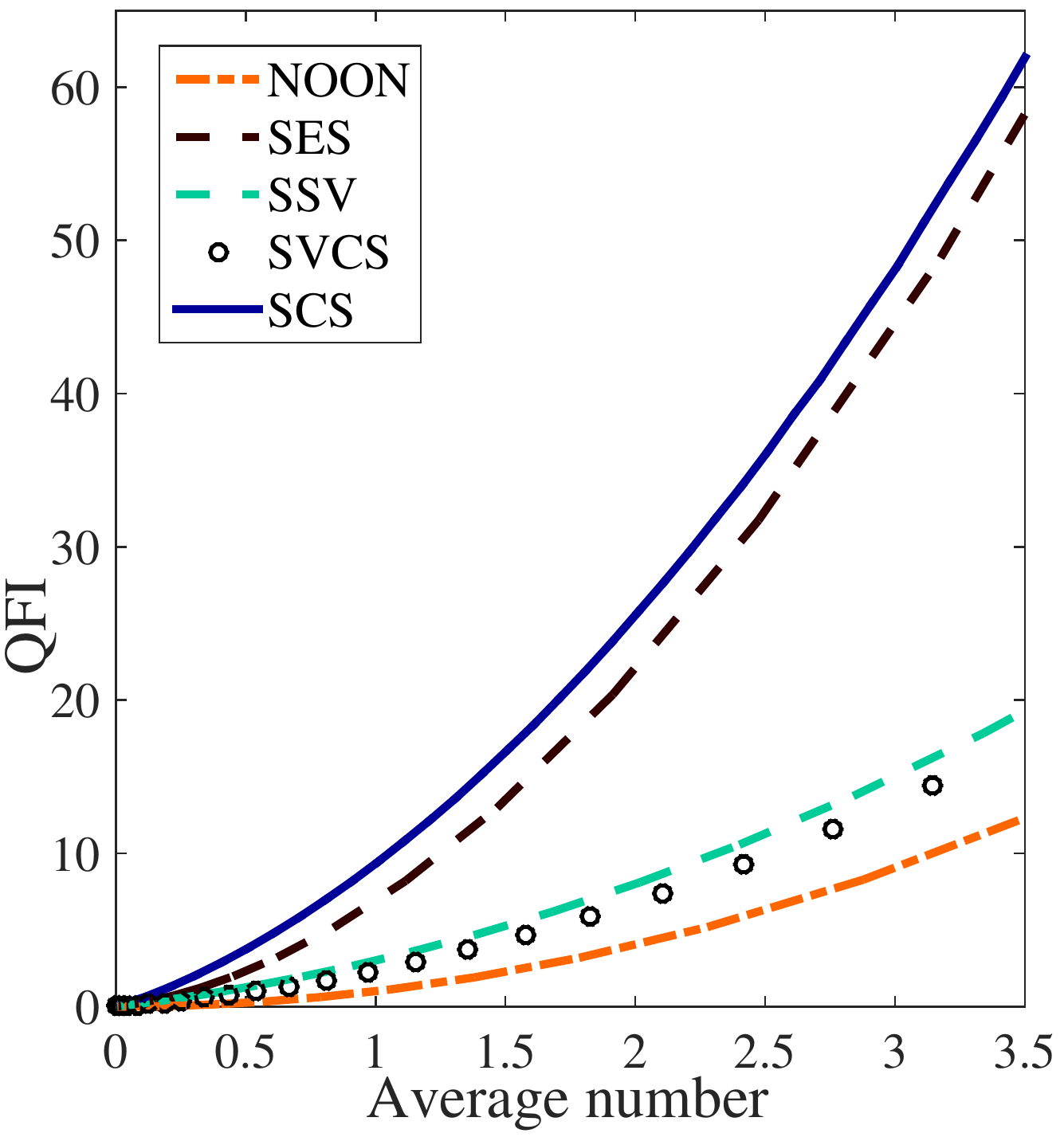}
\caption{The QFI (plotted against average photon number $\bar{n}$) for both the squeezed-entangled state (SES) and the squeezed cat-state (SCS) shows dramatic improvements over the commonly used states for optical quantum metrology, including Caves's state (SVCS), the optimal Gaussian state (SSV), and the NOON state. Furthermore, the squeezed cat-state has been made experimentally \cite{ourjoumtsev2007generation,etesse2015experimental,huang2015optical}, and in this paper we present a measurement scheme that can be employed to read out the phase.}
\label{fig:QFI_SES_vs}
\end{figure}

Remarkably, in the large photon-number limit in which gravitational wave detectors operate, it has been shown that when photon losses are present the original scheme of Caves is optimal \cite{demkowicz2013fundamental}. However, in many applications it is \emph{not} this regime that is of interest and it is instead necessary to consider metrology with low photon numbers. Measurements on fragile systems are of much interest, with examples including measurements of spin ensembles \cite{wolfgramm2013entanglement}, biological systems \cite{carlton2010fast,taylor2013biological}, atoms \cite{tey2008strong,eckert2008quantum} and single molecules \cite{pototschnig2011controlling}, and in all these applications it is of utmost importance to minimize the probe's interaction with the sample to avoid damage. Examples of such damage are the scattering induced depolarisation of spin ensembles \cite{wolfgramm2013entanglement}, or direct degradation of living cells \cite{taylor2015quantum}. It is this small photon number regime that is considered herein, and whilst in this case theoretical lower bounds on precision do exist \cite{demkowicz2015quantum}, it is an open question as to which practical states can give significant improvements over the SNL. In this paper we make significant progress towards this question by introducing an experimentally realisable scheme that can measure to a precision with a $\sqrt{7}$ factor improvement over the SNL, and a $\sqrt{3}$ improvement over the commonly used quantum states, including Caves's scheme \cite{caves1981quantum}.

The general setting of optical quantum metrology can be understood in terms of a two-mode (two-path) interferometer. The enhancement gained from employing quantum states for phase estimation can be then framed in terms of different types of correlations: those between photons on each mode of the interferometer (intra-mode), as well as the correlations between the paths (inter-mode). Both types of correlations can contribute to improvements in precision, and hence it is natural to consider states in which both are present. Observing that the squeezed vacuum exhibits high intra-mode correlations due to non-classical photon statistics \cite{demkowicz2015quantum}, and that inter-mode correlations may be provided by mode-entanglement, this naturally leads us to introduce the `squeezed-entangled state' $|\Psi_{_{\text{SES}}}\rangle \propto |z,0\rangle+|0,z\rangle$ where $\ket{z}$ represents the squeezed vacuum which will be defined below. It will be shown that the fundamental bound on the phase precision possible with this state is a substantial improvement over the states normally considered in the literature, including the state proposed by Caves  \cite{caves1981quantum}, the NOON state \cite{lee2002quantum}, and the optimal Gaussian state (created from only Gaussian transformations) \cite{demkowicz2015quantum,pinel2012ultimate}.

The squeezed-entangled state (SES) has clear potential for precision phase estimation but has the significant disadvantage that it is not clear if a simple high-fidelity preparation procedure can be found. Hence we introduce a practical alternative, the `squeezed cat-state' (SCS), which has been demonstrated experimentally \cite{ourjoumtsev2007generation,etesse2015experimental,huang2015optical}. The quantum Fisher information (QFI)  is a useful and commonly used measure which quantifies the phase precision obtainable using a given probe state, and using this metric the potential for phase estimation of both states proposed herein is shown in Fig.~\ref{fig:QFI_SES_vs} (the requisite QFI formalism will be provided in the next section). Intriguingly, as well as being more practical, the SCS also outperforms the SES, showing that this state is of great interest from both a practical and theoretical perspective. Furthermore, it will be seen that the SCS is robust enough to exhibit a precision advantage with up to $27\%$ photon loss. Finally, it is shown that high-precision phase measurements can be obtained both in the ideal and lossy cases using a photon-number counting measurement.

\section{Correlations in optical metrology}

\begin{figure}[t]
\centering
\includegraphics[scale=1.3]{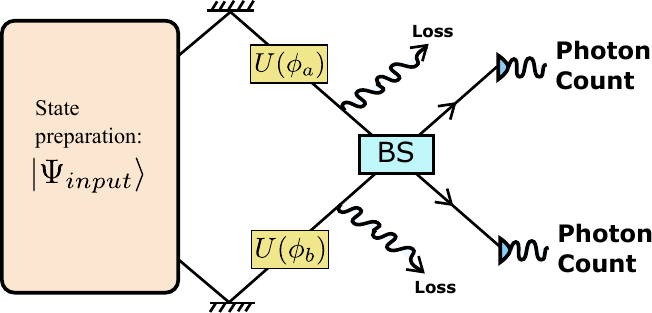}
\caption{A quantum state $\ket{\Psi}$ is prepared as an input into the arms of an interferometer which contains an unknown relative phase shift $\phi \equiv \phi_a - \phi_b$, generated by the linear phase shift unitary operator $\hat{U}=\exp(i (\phi_a \hat{n}_a +\phi_b \hat{n}_b))$. For the states introduced herein the optimal measurement scheme is mixing the modes on a balanced (50:50) beam splitter (BS), followed by photon number counting. When photon losses are considered these can be modelled by `fictitious' variable transmissivity beam splitters after the phase shift.}
\label{fig:TMSC_with_loss}
\end{figure}

We begin by reviewing the relevant background material. In this work we consider the standard optical phase estimation problem of measuring a phase difference $\phi$ between two optical modes containing unknown linear phase shifts, as shown in Fig.~\ref{fig:TMSC_with_loss}. This is applicable to a wide range of physical scenarios and is the canonical approach to a very broad range of metrology schemes. The fundamental limit to the precision with which a state $\rho$ can measure the phase $\phi$ is given by the quantum Cram\'er-Rao bound (CRB) \citep{braunstein1994statistical,braunstein1996generalized}:
\begin{eqnarray}
\Delta \phi \ge \frac{1}{\sqrt{\mu F_Q(\rho)}}, \label{eq:CRB}
\end{eqnarray}
where $\mu$ is the number of independent repeats of the experiment and $F_Q(\rho)$ is the QFI of $\rho$. For pure and path-symmetric states (only path-symmetric states will be considered herein) it is shown in Appendix A that the relevant QFI is simply given by 
\begin{equation} F_Q(\Psi) = 2\left( \text{Var}_{\Psi}  -  \text{Cov}_{\Psi} \right), \label{Eq:FQ1}\end{equation}
where $\text{Var}_{\Psi}= \langle \hat{n}_a^2  \rangle - \langle  \hat{n}_a \rangle ^2$ is the variance of the photon number in mode $a$ (or mode $b$) and $\text{Cov}_{\Psi} =\langle \hat{n}_{a} \otimes \hat{n}_{b} \rangle - \langle  \hat{n}_{a}\rangle\langle \hat{n}_{b} \rangle $ is the covariance of the two modes (the expectation values are taken with respect to the state $\ket{\Psi}$). This explicitly highlights the roles played by inter- and intra-mode correlations.

We now introduce the relevant states in the quantum metrology literature. In the following we denote a coherent state and a squeezed vacuum by $\ket{\alpha} \equiv \hat{D}(\alpha)\ket{0}$ and  $\ket{z} \equiv \hat{S}(z)\ket{0}$ respectively ($\alpha,z \in \mathbb{C}$) where the displacement operator is $\hat{D}(\alpha) = \exp{ (\alpha \hat{a}^{\dagger} - \alpha^* \hat{a}) }$ and the squeezing operator is $\hat{S}(z)=\exp{ \left[ {1 \over 2} (z^* {\hat{a}}^{^2} - z {\hat{a}^{{\dagger}^2}}) \right] }$. Caves \cite{caves1981quantum} proposed the use of squeezing to enhance the phase precision via a probe state obtained from mixing a squeezed vacuum and a coherent state (SVCS) on a balanced (50:50) beam splitter, which is given by $|\Psi_{_{\text{SVCS}}}\rangle =\hat{U}_{_{\text{BS}}} ( |\alpha\rangle_a \otimes |z\rangle_b )$, where $\hat{U}_{_{\text{BS}}}$ denotes the beam splitter unitary operator. This state has been studied extensively and has an asymptotic phase precision of $1/\bar{n}$ \cite{pezze2008mach} (where $\bar{n}=\langle \hat{n}_a + \hat{n}_b \rangle$ is the average number of photons in the interferometer). This precision is known as the Heisenberg limit and is a factor of $1/\sqrt{\bar{n}}$ improvement over the best attainable classical precision given by the shot noise limit (SNL).

An alternative state for quantum-enhanced metrology is the NOON state $|\Psi_{_{\text{NOON}}}\rangle = {1 \over \sqrt{2}} (|N,0\rangle+|0,N\rangle)$ \cite{lee2002quantum} which has a QFI of $F_Q(\Psi_{_{\text{NOON}} })=N^2$ implying a phase precision of $1/N$ \cite{dowling2008quantum}. The NOON state clearly highlights the advantages gained by both inter-mode correlations which are provided by the mode entanglement, and intra-mode correlations which are provided by a large uncertainty in the photon number in each arm. The NOON state is the optimal fixed number state, but if we don't restrict ourselves to fixed number states then improvements over this are possible. We can see this is Fig.~\ref{fig:QFI_SES_vs} where the QFI of the NOON state is plotted against the variable photon-number SVCS. Another variable photon-number state that improves over the NOON state is the separable squeezed vacuum (SSV) given by  $\ket{\Psi_{_{\text{SSV}} }}= \ket{z} \otimes \ket{z}$ which has a QFI of $\bar{n}^2+2\bar{n}$. The SSV is the optimal Gaussian state (a state made with Gaussian operations only), as is described in \cite{pinel2012ultimate,demkowicz2015quantum}. Note that the SSV does not improve over the NOON state in scaling and they have the same precision in the large number limit. We see below that the states introduced in this manuscript obtain factor-improvements over the NOON and SSV even in this limit.
\newline
\newline
\noindent
\section{A squeezed and entangled state}

As discussed in the introduction, exploiting intra- and inter- mode correlations motivates the `squeezed-entangled state' (SES):
\begin{equation}
\label{eq:squent1}
|\Psi_{_{\text{SES}}}\rangle = \mathcal{N} (|z,0\rangle+|0,z\rangle),
\end{equation}
where $\mathcal{N} = (2+2/\cosh|z|)^{-1/2}$. Using equation (\ref{Eq:FQ1}) it can be shown that
\begin{equation}
F_Q(\Psi_{_{\text{SES}} } ) = \frac{3  \bar{n}^2}{2\mathcal{N}^2} + 2\bar{n}, \label{Eq:SESQFI}
\end{equation}
where $\bar{n} = 2\mathcal{N}^2 \sinh^2|z|$. In the large squeezing regime $|z| \gg 1$ we find $\mathcal{N}^2 \approx 1/2$, and hence $F_Q \approx 3\bar{n}^2+2\bar{n}$. This is a factor of 3 better than the NOON state, the SSV and the SVCS in the asymptotic limit, but note that if photon losses are included this asymptotic advantage is lost. In the low photon limit - the regime of interest for this paper - the improvement over the NOON state is even more significant, with $F_Q(\Psi_{_{\text{SES}} } ) \approx 7 F_Q(\Psi_{_{\text{NOON}} } )$ for $\bar{n}=1$. In Fig.~\ref{fig:QFI_SES_vs} we compare the QFI of the SES, the NOON state, the SVCS and the optimal Gaussian state (SSV). Fig.~\ref{fig:QFI_SES_vs} clearly shows the great potential of the SES for quantum enhanced metrology.
\newline
\indent
The SES is a coherent superposition of NOON states of different photon numbers. As NOON states (up to a relative phase factor of $i$) can be generated by inputting $\ket{N} \otimes \ket{0}$ into a non-linear beam splitter \cite{dunningham2006using}, the SES (again up to a relative phase factor of $i$) may similarly be generated in this way via the input of $\ket{z} \otimes \ket{0}$. Alternatively, a method has been proposed that can apply superpositions of squeezing operators in multiple modes \cite{park2015conditional}, which could be used to generate the SES. However, the non-linearities needed for these schemes are not easy to implement physically, and for this reason we look elsewhere for a state that can exploit similar quantum effects to the SES, whilst also being experimentally realisable with current technology.\\

\noindent
\section{The squeezed cat state}

Considering the focus on mode-entanglement in the literature (e.g. NOON states, Holland and Burnett states \cite{holland1993interferometric} and entangled coherent states \cite{gerry2002nonlinear}), it is surprising that inter-mode correlations are not essential for quantum-enhanced metrology \cite{munro2001weak,ralph2002coherent,knott2014effect}. An alternative resource that can be utilized is super-Poissonian photon statistics in the probe state \cite{sahota2015quantum}, which can be seen by writing the QFI of equation (\ref{Eq:FQ1}) in the form
\begin{equation} F_Q = \bar{n} (1 + \mathcal{Q}) (1 - \mathcal{J}) , \label{Eq:FQ2}\end{equation}
where $\mathcal{Q} = (\text{Var}_{\Psi} - \langle \hat{n}_a \rangle) /   \langle \hat{n}_a \rangle $ is the Mandel $\mathcal{Q}$ parameter of mode $a$, and $\mathcal{J} =\text{Cov}_{\Psi}/ \text{Var}_{\Psi}$ \cite{sahota2015quantum}. Interestingly, as pointed out by Sahota and Quesada \cite{sahota2015quantum} $1>\mathcal{J}>-1$, and hence inter-mode correlations (i.e. mode entanglement) can contribute at most a factor of 2 improvement in the QFI ($\mathcal{J}=0$  for a separable state); $\mathcal{Q}$ on the other hand has no upper bound.

In order to find an experimentally viable state with a large Mandel-$\mathcal{Q}$ parameter, a particularly promising avenue of investigation is squeezing a non-Gaussian state. A superposition of coherent states (a cat state) is such a non-Gaussian state, and hence this motivates the introduction of the squeezed cat state (SCS)
\begin{equation} \ket{\psi_{_{\text{SCS}}}} =\mathcal{N} S(z) \left(| \alpha \rangle + |-\alpha \rangle \right) ,\end{equation} 
where $\mathcal{N} = (2+2e^{-2\alpha^2})^{-1/2}$. SCSs may then be used for phase estimation by considering the two-mode state
\begin{equation}
\label{eq:SSCS}
|\Psi_{_{\text{SCS}}}\rangle =\ket{\psi_{_{\text{SCS}}}}_a \otimes \ket{\psi_{_{\text{SCS}}}}_b.
\end{equation}
Clearly this state is mode-separable, although it can be argued it still exhibits entanglement between the photons themselves \cite{demkowicz2015quantum}. The QFI for the SCS as a function of average total photon number (optimized over the parameters $\alpha$ and $z$) is given in Fig.~\ref{fig:QFI_SES_vs}. The analytical formula is presented in Appendix A. The SCS shows a substantial improvement over the SVCS, the NOON state, and the SSV. It even (slightly) improves on the phase precision of the SES introduced above.

The crucial advantage of the SCS over the SES is that the former has been generated experimentally \cite{ourjoumtsev2007generation,etesse2015experimental,huang2015optical}. The method of Ourjoumtsev \textit{et. al.} \cite{ourjoumtsev2007generation} involves splitting a two photon state at a beam splitter, before a projective homodyne measurement is performed on one output mode. An alternative procedure in Ref. \cite{huang2015optical} requires the initial preparation of two squeezed vacuum states. One of the two modes then undergoes a $\pi/2$ phase shift, before the modes are mixed at a beam splitter with variable transmissivity. Finally, a photon number measurement is performed on one mode, heralding the approximate SCS in the remaining mode. With this method Huang \textit{et. al.} have generated an SCS with a fidelity $67\%$ and size $|\alpha|=\sqrt{3}$, making it the largest amplitude coherent state superposition to date \cite{huang2015optical}. Another method for generating an SCS could be to directly squeeze a cat state; there are many examples of cat state generation techniques in the literature \cite{brune1996observing,gerry1997generation,lund2004conditional,bartley2012multiphoton,gerrits2010generation,leghtas2013deterministic}. The subsequent squeezing can be performed in a cavity \cite{guzman2006field,de2004engineering,werlang2008generation}, or by ponderomotive squeezing in an optomechanical system \cite{brooks2012non,safavi2013squeezed,braginski1967ponderomotive}.

\begin{figure}[t]
\centering
\includegraphics[scale=0.6]{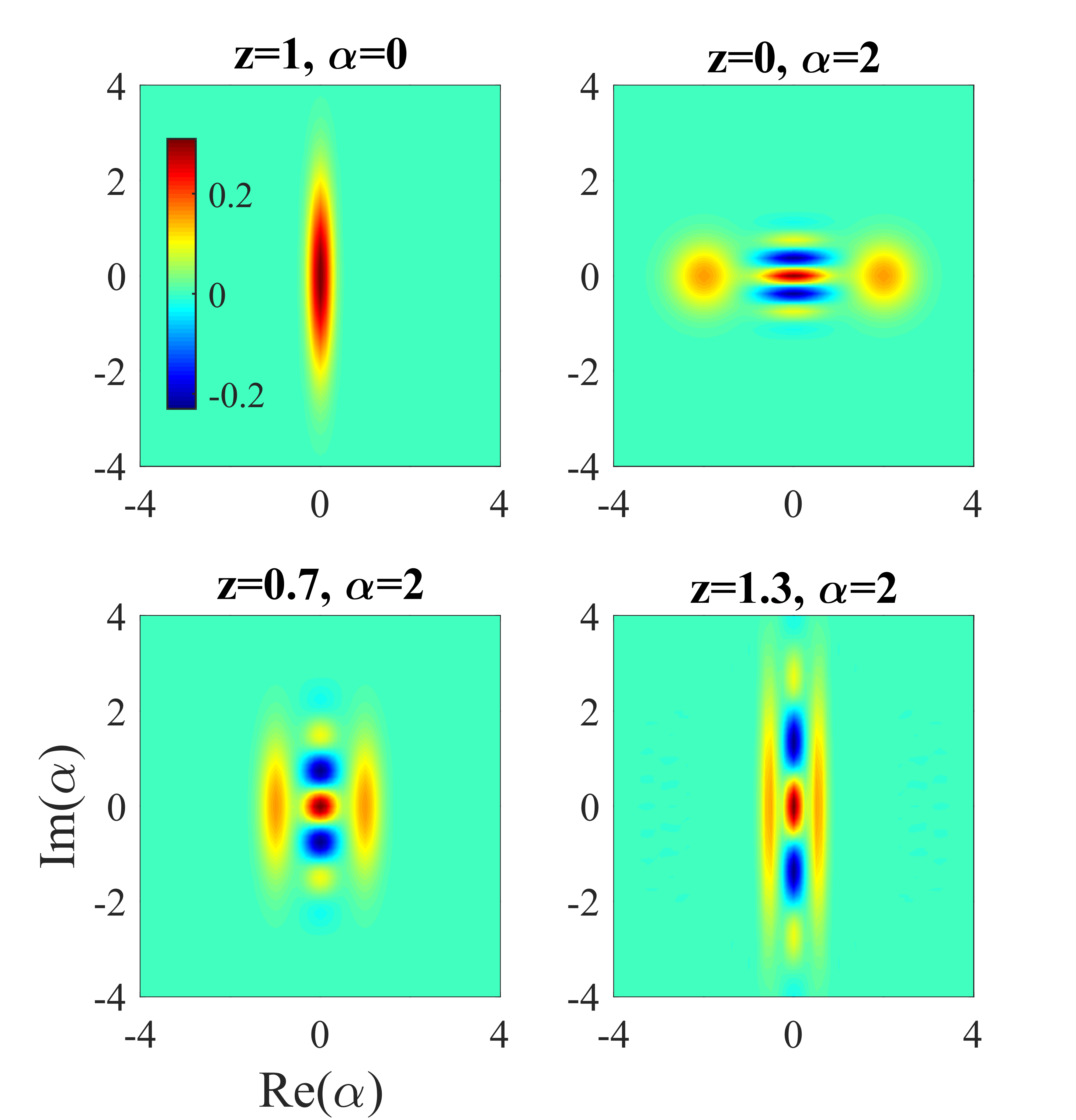}
\caption{We plot the Wigner functions of a squeezed vacuum (top left), a cat state (top right), and the squeezed cat-state (SCS) with different squeezing parameters (bottom row). All axes are as labelled in the bottom left figure. As described in the main text, the QFI is related to the overlap between the Wigner function of a state with and without an infinitesimal phase shift, which is equivalent to a small rotation of the Wigner function about the origin. We see that when the cat state is squeezed, the resultant quasi-probability distribution will exhibit a greater change from a phase rotation, but has a small average photon number. It is clear from the bottom right plot, with $z=1.3$ and $\alpha=2$, that the interference fringes due to the non-Gaussian nature of the state plays a crucial role in minimizing the overlap when the Wigner function is rotated.}
\label{fig:Wigner_func}
\end{figure}

Given the particularly high precision phase estimation possible with the SCS it is interesting to present an intuitive reasoning for these results. A geometric understanding may be obtained by considering the Wigner function of a squeezed cat state, with plots given in Fig.~\ref{fig:Wigner_func}. The link between the Wigner function and the QFI can be made rigorous as follows. The QFI of a pure state can be written in terms of the fidelity, $\mathcal{F}$, between the state $\ket{\Psi}$ and the infinitesimally phase-shifted state $\ket{\Psi(\delta\phi)}$: $F_Q(\Psi) \propto 1-\sqrt{\mathcal{F}_{\Psi}}$, where $\mathcal{F}_{\Psi} \equiv |\langle \Psi(\delta\phi)|\Psi \rangle|^2$\cite{paris2009quantum,banchi2015quantum,taddei2013quantum}. The fidelity can then in turn be written in terms of the overlap of the Wigner functions \cite{caves2004fidelity}:
\begin{equation}
\mathcal{F}_{\Psi}(\phi) = \pi \int d^2\alpha W_{\Psi} W_{\Psi(\delta\phi)}.
\end{equation}
Therefore states for which the overlap of the Wigner functions with and without a phase shift is small exhibit a large QFI. Given that the resource of interest in quantum metrology is the average photon number, the desired property for low-photon high-precision phase estimation is a large change in the Wigner function when rotated about the origin in conjunction with a low average photon number in the state. Fig.~\ref{fig:Wigner_func} indicates that rotating the Wigner function of the SCS results in a small overlap, and therefore a large QFI, whilst retaining small photon numbers.
\newline
\newline
\noindent
\section{The measurement scheme}

The measurement scheme we propose is to mix the modes on a balanced beam splitter followed by photon number counting, as shown in Fig.~\ref{fig:TMSC_with_loss}. This can be challenging, but photon number resolving detectors are an area of intense research \cite{Eisaman2001} and devices that are highly sensitive in the low photon regime, the area most relevant for this work, have been demonstrated \cite{gerrits2010generation,calkins2013high,Fukuda:11, divochiy2008superconducting}. In particular, recent results by Humphreys \textit{et. al.} use a transition-edge sensor to resolve up to 14 photons with over $60\%$ confidence \cite{humphreys2015tomography}. Many schemes like ours will benefit as advances continue to be made with this technology. To assess our measurement scheme we use the classical Fisher information (CFI), which provides the absolute bound on the phase precision obtainable with a specific measurement, and is calculated from the associated probability distribution of measurement outcomes. In our case we obtain the probability distribution $P(m,n)$ of detecting $m$ ($n$) photons at the first (second) output of the beam splitter. The CFI is then given by:

\begin{equation}
\label{eq:CFI}
F_C(\phi) = \sum_{m=0}^{\infty}\sum_{n=0}^{\infty} {1 \over P(m,n)} \left( \frac{\partial P(m,n)}{\partial \phi} \right)^2.
\end{equation}
Using this we find that our measurement scheme saturates the bound given by the QFI for the majority of $\phi$ values. Indeed, this is to be expected as such a measurement is optimal for any pure and path-symmetric state \cite{hofmann2009all}. As with most quantum metrology schemes we can't saturate the QFI for every phase $\phi$, and therefore if a completely unknown phase is being measured then an adaptive strategy should be used \cite{xiang2011entanglement,berry2000optimal}. The fact that the measurement scheme saturates the bound confirms that there is approximately a factor of $\sqrt{3}$ improvement in the phase estimation provided by the SCS over the optimal Gaussian state. To highlight the importance of this result, we note that the optimal Gaussian state can improve over, or equal, all of the quantum metrology states in recent experiments (known to the authors). This includes the squeezed states, which have been used in gravitational wave detectors \cite{aasi2013enhanced}, biological sensing \cite{taylor2013biological}, spin noise spectroscopy \cite{lucivero2015squeezed} and the ultrasensitive measurement of a microcantilever displacement \cite{pooser2015ultrasensitive}.

It is important to now address some limitations inherent in using the QFI and CFI as figures of merit in quantum metrology. In general, the precisions as obtained by the QFI and CFI are achievable with an asymptotically large number of repeats, $\mu$. However, from a practical point of view it is clear that only some finite number of repeats will be possible (this may be limited by the fragility of the physical system). The experimenter's prior knowledge of the phase also has to be considered in any realistic setting. Indeed, states with unbounded QFI for fixed $\bar{n}$ can be found  \cite{zhang2013unbounded,rivas2012sub}, but when the required repeats or prior information are considered it has been shown that such states cannot provide a sub-Heisenberg scaling \cite{hall2012universality,hall2012heisenberg,giovannetti2012sub}.

To mitigate the potential problems that can arise when using the QFI we have therefore performed a Bayesian simulation of the proposed experiment. This properly accounts for the information that would be obtained in an experiment rather than relying on the bound given by the QFI. Using the measurement scheme in Fig.~\ref{fig:TMSC_with_loss} we have determined the phase shift, from a flat prior knowledge, using the Bayesian approach described in Appendix B. The simulations confirm that we come close to saturating the absolute bound given by the QFI and equation (\ref{eq:CRB}) for $\mu=O(10^2)$. In such regimes it is then clear that the SCS and the SES can significantly outperform the alternative states in terms of absolute phase precision, when assuming the same average photon number. Note that this is \emph{not} claiming a sub-Heisenberg scaling; from a practical perspective, the scaling with $\bar{n}$ is not necessarily the most relevant quantity as only small quantum states are likely to be available. Indeed the scaling is of no relevance when considering fragile systems which can tolerate a specified (approximate) maximum number of photons.\\

\noindent
\section{The effects of loss}

We next investigate the effects of loss on the squeezed cat-state, which can be modelled by adding `fictitious' beam splitters after the phase shift \cite{leonhardt1993realistic,demkowicz2009quantum} as shown in Fig.~\ref{fig:TMSC_with_loss}. Loss destroys quantum effects, and hence any non-classical enhancement will be reduced when loss is considered. The QFI for a general density matrix $\rho$ can be expressed as \cite{braunstein1994statistical,braunstein1996generalized,joo2011quantum}:
\begin{equation}
F_Q = \sum_{i,j} \frac{2}{\lambda_i+\lambda_j}\left| \langle \lambda_i | \partial \rho(\phi) / \partial \phi| \lambda_j \rangle\right|^2,
\label{QFI}
\end{equation}
where $\lambda_i$ are the eigenvalues and $|\lambda_i \rangle$ a corresponding set of orthonormal eigenvectors of $\rho$.

\begin{figure}[t]
\centering
\includegraphics[scale=0.47]{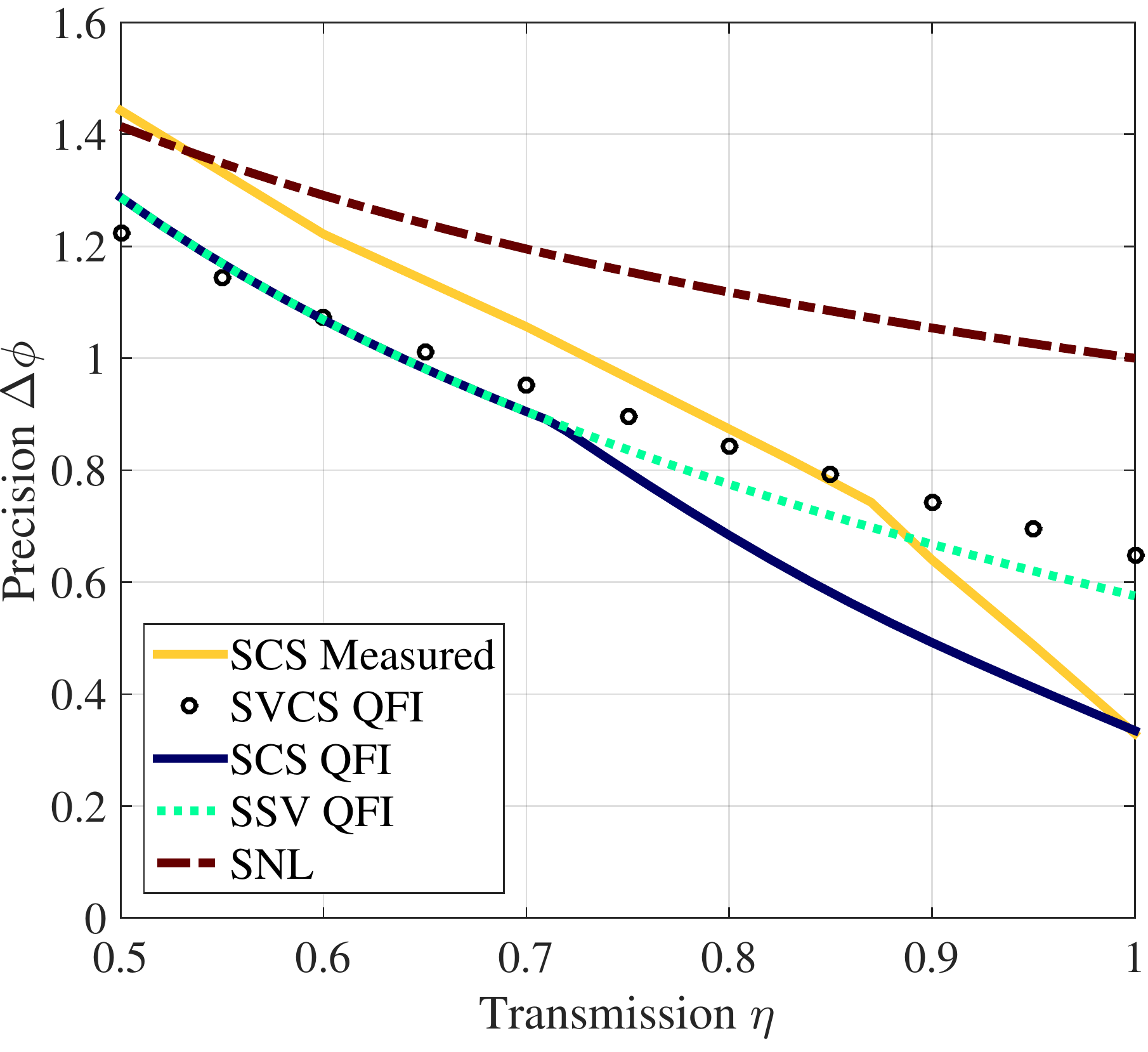}
\caption{The transmission probability through the interferometer, $\eta$, is plotted against the precision, $\Delta\phi$ (scaled by $\sqrt{\mu}$), for various states. The precision is found from equation (\ref{eq:CRB}) ($F_Q$ is replaced with $F_C$ for the `SCS Measured' curve). The QFI of the SCS demonstrates the potential for robust phase measurements up to $27\%$ loss. The measurement scheme in Fig.~\ref{fig:TMSC_with_loss} is then plotted for the SCS, showing that with a conceptually simple scheme, without external reference beams, the SCS can still beat the alternative states up to $10\%$ loss.}
\label{fig:Loss_QFI_CFI_final1}
\end{figure}

The precision (given by the QFI and equation (\ref{eq:CRB})) as a function of loss is plotted in Fig.~\ref{fig:Loss_QFI_CFI_final1}, optimized over the state parameters $z$ and $\alpha$, whilst fixing the average number of photons in our state at $\bar{n} = 1$. For low loss and low photon numbers, the improvement is a factor of $\sqrt{7}$ greater than the SNL (which is identical to the NOON state for $\bar{n} = 1$). The SCS is robust enough to exhibit a precision advantage up to $27\%$ loss. Fig.~\ref{fig:Loss_QFI_CFI_final1} also shows the results of our measurement scheme, calculated from the CFI in equation (\ref{eq:CFI}) substituted into equation (\ref{eq:CRB}) ($F_Q$ is replaced with $F_C$). We see that the SCS improves over the best possible measurement, as given by the QFI, of both the optimal Gaussian state (SSV) and the SVCS, for losses up to $10\%$. Losses as low as $10\%$ have already been achieved in table top interferometry experiments \cite{eberle2010quantum}, and near-future gravitational wave detectors are expected to have total losses of $9-17\%$ \cite{oelker2014squeezed}. We note that a major advantage of the phase readout presented here is that the measurement scheme does not have to be altered when loss is present, for example by using extra reference beams, as in \cite{knott2014attaining} or \cite{ono2010effects}.\\

\noindent
\section{Conclusion}

In this paper we have introduced quantum states that exhibit large factor improvements in the phase-estimation precision over the commonly used states for quantum metrology. Motivated by considering both inter- and intra- mode correlations we introduced the squeezed-entangled state (SES), which demonstrates a 7-fold enhancement in the quantum Fisher information over the NOON state and a 3-fold improvement over the optimal Gaussian state, for low photon numbers. The question of practicality was then addressed by introducing the squeezed cat-state (SCS), which has been experimentally generated \cite{ourjoumtsev2007generation,huang2015optical,etesse2015experimental}, and exhibits even greater enhancements in the attainable phase precision than the SES. A conceptually simple measurement scheme that saturates the theoretical phase-precision bound when there is no loss was given, and the robustness of the SCS to moderate loss for small photon numbers was demonstrated. These results illustrate that substantial precision improvements can be made over the quantum states traditionally proposed for practical optical metrology, and as the SCS has already been generated we expect that an experiment could confirm our results in the near future.

\section{acknowledgements}

This work was partly funded by the UK EPSRC through the Quantum Technology Hub: Networked Quantum Information Technology (grant reference EP/M013243/1), and by UK Ministry of Defence through DSTL's National UK PhD programme (contract number DSTLX1000095642). After completing this work we became aware of related results on the SES \cite{lee2015quantum}.


\noindent
\section{Appendix A: Quantum Fisher information}
The definition of the QFI for an arbitrary mixed state $\rho_{\phi}$ which depends on a single parameter $\phi$ is $F_Q(\rho_{\phi})= \text{tr} ( \rho_{\phi} L(\rho_{\phi})^2 )$ where $L(\rho_{\phi})$ is the \emph{symmetric logarithmic derivative} defined implicitly by $\frac{\partial}{\partial \phi} \rho_{\phi} = \rho_{\phi} L(\rho_{\phi}) +  L(\rho_{\phi})\rho_{\phi}  $ \cite{demkowicz2015quantum}. The mixed state QFI formula given in equation ~(\ref{QFI}) and used in the case of photon losses may be derived from this definition \cite{braunstein1994statistical,braunstein1996generalized,joo2011quantum}. For a pure state $\ket{ \psi_{\phi} } $, the QFI reduces to
\begin{equation} F_Q(\psi_{\phi} ) = 4 \left( \langle \partial_{\phi} \psi_{\phi} | \partial_{\phi} \psi_{\phi} \rangle - | \langle \partial_{\phi} \psi_{\phi} |  \psi_{\phi} \rangle |^2  \right), \end{equation}
with $\ket{\partial_{\phi} \psi_{\phi}}  \equiv \frac{\partial}{\partial \phi} \ket{\psi_{\phi}}$. It may be confirmed with simple algebra that if the parameter is imprinted on the quantum state by a unitary transformation of the form
 $\hat{U}(\phi) = \exp ( i \phi \hat{O} )$, then the QFI is proportional to the variance of the generating operator, specifically $F_Q =  4 \text{Var}_{\Psi} (  \hat{O} )$ where $\text{Var}_{\Psi} (  \hat{O} )= \langle \hat{O}^2  \rangle - \langle  \hat{O} \rangle ^2$.

The phase-estimation problem under consideration herein is summarized in Fig.~\ref{fig:TMSC_with_loss}.
A two-mode quantum state $\ket{\Psi}$ undergoes unknown linear phase shifts in each mode, i.e. it evolves via the unitary operator $U\equiv \exp(i (\phi_a \hat{n}_a +\phi_b \hat{n}_b))= \exp(i (\phi^+ \hat{O}^+ +\phi^- \hat{O}^-))$ for unknown $\phi_a$ and $\phi_b$, where  $\hat{O}^{\pm}=(\hat{n}_a \pm \hat{n}_b)/2$ and $\phi^{\pm} = \phi_a \pm \phi_b$. The aim is to estimate the relative phase shift $\phi \equiv \phi^-= \phi_a - \phi_b$. If a phase reference (with respect to which each phase is defined) is available then this is a two-parameter estimation problem ($\phi^{\pm}$), which requires a two-parameter QFI \cite{jarzyna2012quantum}, and if a phase reference is not available the total phase ($\phi^+$) is of no physical relevance and this should be averaged over, creating a mixed state. In this case it is in general necessary to use the mixed state QFI. However, when the input is \emph{path-symmetric and pure}, it has been shown that the phase averaging has no effect on the QFI \cite{jarzyna2012quantum} and the relevant QFI formula reduces to $F_Q(\Psi) =  \text{Var}_{\Psi}(\hat{n}_a - \hat{n}_b)$, as simply obtained by using $F_Q(\Psi) =  4  \text{Var}_{\Psi} (  \hat{O} )$ with the generator for the phase shift $\hat{O}^{-}$. As only path-symmetric states are considered herein, this simple QFI formula may be used in the lossless case. However, the use of this QFI in cases without path-symmetry can lead to over-optimistic bounds on the phase precision as explained in detail in Ref. \cite{jarzyna2012quantum}. By expanding this variance it then follows that $F_Q(\Psi) = 2\left( \text{Var}_{\Psi}  -  \text{Cov}_{\Psi} \right)$, as stated in equation (\ref{Eq:FQ1}) and using the notation defined there. Clearly due to path-symmetry the single-mode variance may be calculated with respect to either mode. To obtain equation (\ref{Eq:FQ2}) from this formula requires only basic algebra \cite{sahota2015quantum}. For all path-symmetric pure states, the optimal measurement scheme which saturates the QFI, is mixing the modes on a balanced beam splitter, followed by photon number counting \cite{hofmann2009all}.
\newline
\indent
The QFI of the two-mode squeezed cat state may be calculated from equation (\ref{Eq:FQ1}). As the state is separable $\text{Cov}_{\Psi} =0$ and hence $F_Q(\Psi_{_{\text{SCS}}}) = 2 \text{Var}_{\Psi}$, which depends only on the variance in the photon number in a single-mode squeezed cat state $\ket{\psi_{_{\text{SCS}}}}$. A direct calculation of this quantity yields
\begin{multline}
 F_Q(\Psi_{_{\text{SCS}}})=4(s^4_1+s^2_1)+2\alpha^2(\tau c_4-s_4) \\+2\alpha^4 \left(c_4- \tau s_4 -(\tau c_2- s_2)^2\right) \end{multline}
where $s_k \equiv \sinh(kz)$, $c_k \equiv \cosh(kz)$ and $\tau =  (2-2e^{-2\alpha^2}) (2+2e^{-2\alpha^2})^{-1}$. The average total number of photons in the (two-mode) state is 
\begin{equation}\bar{n} = 2s^2_1 + 2\alpha^2(\tau c_2 -s_2).\end{equation}
Note that as this state contains two parameters the QFI may not in general be expressed directly in terms of $\bar{n}$ only. As the aim is to maximize the phase precision for a given average number of photons, the optimal choice of parameters $\alpha$ and $z$ for each $\bar{n}$ is found by maximizing the QFI for each fixed average particle number. Note that the special case of $\alpha=0$ results in $F_Q=\bar{n}^2 + 2\bar{n}$ and hence this provides a lower bound on the optimized QFI. The maximisation over $\alpha$ and $z$ was performed numerically and it is this resultant function that is plotted in Fig.~\ref{fig:QFI_SES_vs}.
\newline
\noindent
\section{Appendix B: Bayesian simulation}
The advantage of performing a Bayesian simulation is that it replicates how data is gathered in an experiment. It therefore gives a reliable measure of the precision that can be achieved, rather than relying on the QFI and CFI bounds which, as noted in the main text, can be misleading.

To implement the Bayesian simulation for the scheme in  Fig.~\ref{fig:TMSC_with_loss} we first calculate the probability of detecting $(m,n)$ photons at the two output ports: $P(m,n|\phi)$. The simulation begins by selecting a phase $\phi_0$ that an experimenter wishes to measure (the experimenter does \emph{not} have access to the value of $\phi_0$). We can then calculate the conditional probability of detecting $(m,n)$ particles at the output ports, \textit{given} that the phase is $\phi_0$: $P(m,n|\phi_0)$. A random outcome is sampled from this distribution which gives a pair of values $(m_1,n_1)$: these are the simulated outputs which correspond to what the experimenter measures after they send the given state through the interferometer.
\newline
\indent
The experimenter must now try to determine the phase from their measured values $(m_1,n_1)$. To do this, they use Bayes' theorem: $P(a|b) \propto P(b|a)$. The experimenter can then calculate:
\begin{eqnarray}
P(\phi|m_1,n_1) \propto P(m_1,n_1|\phi).
\end{eqnarray}
As the probability distribution sums to one, they can normalize this distribution to be left with $P(\phi|m_1,n_1)$: the probability distribution for different phases $\phi$ \textit{given} that $(m_1,n_1)$ has been measured. In our simulation we repeats these steps, allowing the experimenter to gain more knowledge about the phase. With each new measurement the experimenter can use Bayesian inference to update their knowledge of the phase.
\newline
\indent
After a number of repeats, the experimenter is left with a probability distribution $P(\phi)$, which is the probability distribution for $\phi$, \textit{given} all previous measurements at the detectors. The precision with which we can measure the phase is then taken to be the standard deviation of this probability distribution. Taking an average over many simulations provides the $\Delta\phi$ results that are described in the main text.


\section{References}

\bibliographystyle{apsrev}
\bibliography{MyLibrary_Thesis_plus_squent_Apr2015}

\end{document}